\documentstyle [twocolumn,prl,aps]{revtex}
\begin{document}
\twocolumn [\hsize\textwidth\columnwidth\hsize\csname
@twocolumnfalse\endcsname

\title{From spatially indirect to momentum-space indirect
exciton by in-plane magnetic field}

\author{L. V. Butov$^{1}$, A. V. Mintsev$^1$, Yu.E. Lozovik$^2$,
K. L. Campman$^3$, and A. C. Gossard$^3$}

\address{$^1$Institute of Solid State Physics, Russian Academy
of Sciences, 142432 Chernogolovka, Russia\\
$^2$Institute of Spectroscopy, Russian Academy of Sciences,
142092, Troitsk, Russia\\
$^3$Department of Electrical and Computer Engineering and Center
for Quantized Electronic Structures (QUEST), University of
California, Santa Barbara, CA 93106}

\date{December 11, 1999}
\maketitle
\begin{abstract}
In-plane magnetic field is found to change drastically the
photoluminescence spectra and kinetics of interwell excitons in
GaAs/Al$_x$Ga$_{1-x}$As coupled quantum wells. The effect is due
to the in-plane magnetic field induced displacement of the
interwell exciton dispersion in a momentum space, which results
in the transition from the momentum-space direct exciton ground
state to the momentum-space indirect exciton ground state.
In-plane magnetic field is, therefore, an effective tool for the
exciton dispersion engineering.

\end{abstract}

\vskip 2pc ] % end \twocolumn[...]
\narrowtext

The effective exciton temperature in a quasiequilibrium system
of excitons in semiconductors is determined by the ratio between
the exciton energy relaxation rate and the exciton recombination
rate. The low exciton temperature is crucial for an observation
of a number of novel collective phenomena caused by the high
occupation of the lowest energy exciton states in a
quasi-two-dimensional (2D) exciton system
\cite{Lozovik7,Kuramoto,Yoshioka,Chen,Zhu}. Low temperatures can
be achieved in a system with a low exciton recombination rate.

A long exciton lifetime is characteristic (1) for the systems
where the ground state exciton is optically inactive (in dipole
approximation) because of parity, e.g. for Cu$_2$O; (2) for the
systems where electron and hole are spatially separated, e.g.
for spatially indirect (interwell) excitons in direct-band-gap
coupled quantum wells (CQWs) like $\Gamma - X_z$ AlAs/GaAs CQWs
and GaAs/Al$_x$Ga$_{1-x}$As CQWs; (3) for the systems where the
ground state exciton is indirect in a momentum space, e.g. for
$\Gamma - X_{xy}$ AlAs/GaAs CQWs.

Due to the coupling between the internal structure of
magnetoexciton and its center-of-mass motion \cite{Gor'kov} the
ground exciton state in a direct-band-gap semiconductor in
crossed electric and magnetic fields was predicted to be at
finite momentum. In particular, the transition from the
momentum-space direct exciton ground state to the momentum-space
indirect exciton ground state was predicted (1) for interwell
exciton in coupled quantum wells in in-plane magnetic field
\cite{Lozovik7,GT}, and (2) for single layer exciton in in-plane
electric field and perpendicular magnetic field
\cite{Dzyubenko,Paquet,Imamoglu}. These effects should allow for
a controllable variation of the exciton dispersion and increase
of the exciton ground state lifetime.

In the present paper we report on the experimental observation
of the in-plane magnetic field induced transition from the
momentum-space direct exciton ground state to the momentum-space
indirect exciton ground state for the system of interwell
excitons in GaAs/Al$_x$Ga$_{1-x}$As CQWs. The transition is
identified by the drastic change of the exciton
photoluminescence (PL) kinetics. The interwell exciton
dispersion in in-plane magnetic field is analyzed theoretically
and determined experimentally from the shift of the interwell
exciton PL energy; the experimental data are found to be in a
qualitative agreement with the theory.

We suppose that the in-plane magnetic field is directed along
the $x$-axis, and use the calibration $A_x=A_z=0, A_y=-Bz$. Due
to the invariance corresponding to the simultaneous magnetic
translation of electron and hole on the same vector parallel to
the CQW plane, i.e. the invariance to the translation and
corresponding gauge transformation, the two-dimensional magnetic
momentum is conserved \cite{Gor'kov}. For the gauge used
\begin{equation}
P_x=p_{ex}+p_{hx}; P_y=p_{ey}+p_{hy}-{e\over{c}}B(z_e-z_h).
\label{eq:1}
\end{equation}
For $z^{\prime}_{e,h}$ measured from the centers of the
corresponding QW's one obtains
$P_y=p_{ey}+p_{hy}-{e\over{c}}Bd-{e\over{c}}B(z^{\prime}_e-z^{\prime}_h)$,
where $d$ is the mean separation between the electron and hole
layers. The value $p_B=-{e\over{c}}Bd=\hbar d/l_B^2$ is the
shift of the magnetic momentum of interwell exciton in the
ground state (as follows from the analysis of Shrodinger
equation); $l_B=({eB\over{\hbar c}})^{-{1\over{2}}}$ is the
magnetic length.

The physical sense of this shift can be obtained from the
analysis of the adiabatic turning on of the in-plane magnetic
field. The appearance of the vortex electric field leads to the
acceleration of interwell exciton. The final in-plane momenta is
equal precisely to $-{e\over{c}}Bd$ and directed normal to the
magnetic field. Therefore, the appearance of the momentum
$-{e\over{c}}Bd$ is related to the diamagnetic response of
electron-hole (exciton) system in CQW in in-plane magnetic
field.

The contribution of the second order on the magnetic field
consist of two parts: the first, depending on the momenta, is
the sum of Van Vleck paramagnetism of isolated QWs, and the
second is the sum of diamagnetic shifts of isolated QWs
\cite{Stern}. Van Vleck paramagnetism leads to the
renormalization of the effective (magnetic) mass of exciton
along the $y$-axis, $M_{yy}$. Therefore, the magnetoexciton
dispersion law becomes anisotropic:
\begin{equation}
M_{xx}=M=m_e+m_h; M_{yy}=M+\delta M(B),
\label{eq:2}
\end{equation}
where
\begin{equation}
\delta M(B)={e^2B^2\over{c^2}}(f_e+f_h),
\label{eq:3}
\end{equation}
\begin{equation}
-f_{e,h}=\sum_{n}{
{\left\vert <0 \left\vert z_{e,h} \right\vert n>\right\vert
^2}\over{E_0-E_n}}.
\label{eq:4}
\end{equation}
Here $E_n, \vert n>$ are the energies and state vectors
corresponding to the size quantization of QW's; $f_{e,h}$ are
related to the electric polarization of QW's in the
$z$-direction. $\delta M(B)>0.$ The estimate results to:
\begin{equation}
{\delta M_{yy}\over{M}} \sim {E_{diam}\over E} \sim
{\left({L_z\over{l_B}}\right)^4},
\label{eq:5}
\end{equation}
where $E_{diam}$ is the diamagnetic shift in isolated QW, $E$ is
the size quantized excitation energy of QW, $L_z$ is the QW
width.

The electric-field-tunable $n^+-i-n^+$ GaAs/AlGaAs CQW structure
was grown by MBE. The $i$-region consists of two 8~nm GaAs QWs
separated by a 4~nm Al$_{0.33}$Ga$_{0.67}$As barrier and
surrounded by two 200~nm Al$_{0.33}$Ga$_{0.67}$As barrier
layers. The $n^+$-layers are Si-doped GaAs with $N_{\rm
Si}=5\times 10^{17}$~cm$^{-3}$. The electric field in the
z-direction is monitored by the external gate voltage $V_g$
applied between $n^+$-layers. Carriers were photoexcited by a
pulsed semiconductor laser ($\hbar\omega =1.85$~eV, the pulse
duration was about 50~ns, the edge sharpness including the
system resolution was $\approx 0.5$~ns, the repetition frequency
was 1~MHz, $W_{ex}=10$ W/cm$^2$). In order to minimize the
effect of the mesa heating, we worked with a mesa area of
$0.2\times 0.2$~mm$^2$, which was much smaller than the sample
area of about 4~mm$^2$. In addition, the bottom of the sample
was soldered to a metal plate. The PL measurements were
performed in a He cryostat by means of an optical fiber with
diameter 0.1 mm positioned 0.3 mm above the mesa. The PL spectra
and kinetics were measured by a time correlated photon counting
system.

The separation of electrons and holes in different QWs (the
indirect regime) is realized by applying a finite gate voltage
which fixes external electric field in the $z$-direction
$F=V_g/d_0$, where $d_0$ is the $i$-layer width \cite{CQW}. The
spatially direct (intrawell) and indirect (interwell)
transitions are identified by the PL kinetics and gate voltage
dependence: the intrawell PL line has short PL decay time and
its position practically does not depend on $V_g$, while
interwell PL line has long PL decay time and shifts to lower
energies with increasing $V_g$ (the shift magnitude is given by
$eFd$) \cite{CQW,CQW1}. The upper and the lower direct
transitions are related to the intrawell 1s heavy hole exciton
$X$ and intrawell charged complexes $X^+$ and $X^-$
\cite{trion}.

With increasing in-plane magnetic field the energy of interwell
exciton increases while the energies of the direct transitions
are almost unaffected (Fig.\ 1a). This behaviour is consistent
with a displacement of the interwell exciton dispersion in
in-plane magnetic field. The scheme of the interwell exciton
dispersion at zero and finite in-plane magnetic field is shown
in Fig.\ 2.

For delocalized 2D excitons only the states with small
center-of-mass momenta $k \le k_0 \approx E_g / \hbar c$ (where
$c$ is the speed of light in the medium), i.e. within the
radiative zone, can decay radiatively \cite{Feldmann} (see Fig.\
2). For GaAs structures $k_0 \approx 3 \times 10^5$ cm$^{-1}$
and is much smaller than $k_B=p_B/\hbar$ in strong fields (at
$B=10$ T $k_B \approx 2 \times 10^6$ cm$^{-1}$). The energy of
the interwell exciton PL is set by the energy of the radiative
zone.  Therefore, as the diamagnetic shift of the bottom of the
subbands is small and can be neglected to the first
approximation, the interwell exciton PL energy in in-plane
magnetic field should be increased by $E_{p=0} =
p_B^2/2M=e^2d^2B^2/2Mc^2$. In particular, the in-plane magnetic
field dependence of the interwell PL energy could be used for
the measurement of the exciton dispersion because it allows to
access high momentum exciton states. At small fields the
measured PL energy shift rate is 0.062~meV/T$^2$ (see inset to
Fig.\ 1a) which corresponds to $M=0.21 m_0$. This value is close
to the calculated mass of heavy hole exciton in GaAs QWs
$\approx 0.25 m_0$ ($m_e=0.067 m_0$ and the calculated in-plane
heavy hole mass near $k=0$ is reported to be $m_h=0.18~m_0$, see
Ref. \cite{Bauer} and references therein). Figure 1 shows
considerable deviation from the quadratic dependence of the
interwell exciton PL energy at high $B$. This deviation is the
consequence of the interwell exciton mass renormalization due to
in-plane magnetic field which is predicted by theory, see Eqs.\
2-4. Due to the estimate of Eq.\ 5 the deviation of the
interwell exciton dispersion from the quadratic one should
become essential in the magnetic field where $l_B \sim L_z$.
This is in a good qualitative agreement with the experiment
which presents the onset of the deviation at $B \approx 8$ T,
where $l_B=9$ nm $\sim L_z=8$ nm (see inset to Fig.\ 1a). A
possible contribution from the hole dispersion nonparabolicity
to the observed increase of the exciton mass is, apparently, a
minor effect for the small exciton energies considered which are
much smaller than the light-\ heavy-hole splitting equal to 17
meV for the CQW studied \cite{trion}.

Note that the development of indirect gap when an in-plane
magnetic field is applied has been observed in asymmetric
modulation doped single quantum well where the centers of the
electron and hole envelopes do not coincide; due to the free
carrier character of recombination in the studied 2D electron
gas, the PL energy shift corresponded to the electron mass
\cite{Whittaker}.

In-plane magnetic field modifies qualitatively the exciton PL
kinetics (Fig.\ 1b). The main feature of the interwell exciton
PL kinetics at zero magnetic field is a sharp enhancement of the
PL intensity after the excitation is switched off - the PL-jump
\cite{CQW1} (Fig.\ 1). The basis of the effect is the following.
The exciton PL kinetics is determined by the kinetics of
occupation of the radiative zone (marked bold in Fig. 2). The
occupation varies due to the exciton recombination and energy
relaxation. The PL-jump denotes a sharp increase of the
occupation of the optically active exciton states just after the
excitation is switched off. It is induced by the sharp reduction
of the effective exciton temperature, $T_{eff}$, due to the fast
decay of the nonequilibrium phonon occupation and energy
relaxation of hot photoexcited excitons, electrons, and holes
\cite{CQW1}. The disappearance of the PL-jump at high in-plane
magnetic fields is consistent with the displacement of the
interwell exciton dispersion in parallel magnetic field: for
momentum-space indirect exciton a sharp reduction of $T_{eff}$
just after the excitation is switched off should not result in
the increase of occupation of the radiative zone (see Fig.\ 2c)
and, hence, the PL intensity.

The measured radiative decay rate is proportional to the
fraction of excitons in the radiative zone. The observed strong
reduction of the radiative decay rate in in-plane magnetic field
(by more than 20 times in $B=12$ T, see Fig.\ 1b) also reflects
the displacement of the interwell exciton dispersion and,
correspondingly, nonradiative character of the ground exciton
state in parallel magnetic field. In high in-plane magnetic
field the radiative decay rate becomes comparable and smaller
than the nonradiative decay rate which results to the observed
quenching of the interwell exciton PL intensity (Fig.\ 1a).

The unambiguous evidence for the nonradiative character of the
ground state of interwell exciton in parallel magnetic field has
been observed from the temperature dependence of the PL kinetics
(Fig.\ 3). At zero field the exciton recombination rate
monotonically {\it reduces} with increasing temperature (Fig.\
3b) due to the thermal reduction of the radiative zone
occupation \cite{CQW1}. In high in-plane magnetic field the
temperature dependence is opposite: the exciton recombination
rate $\it enhances$ with increasing temperature (Fig.\ 3a) due
to the increasing occupation of the radiative zone.
Correspondingly, with increasing temperature the interwell
exciton PL intensity reduces at zero field and enhances in high
in-plane magnetic field (see insets to Fig.\ 3).

In conclusion, we have observed drastic change of the
photoluminescence spectra and kinetics of interwell excitons in
GaAs/Al$_x$Ga$_{1-x}$As CQWs in in-plane magnetic field. The
effect is due to the in-plane magnetic field induced
displacement of the indirect exciton dispersion in a momentum
space, which results in the transition from the momentum-space
direct exciton ground state to the momentum-space indirect
exciton ground state.  In-plane magnetic field is, therefore, an
effective tool for the exciton dispersion engineering and
controllable increase of the exciton ground state lifetime. We
speculate that it can be used for the experimental realization
of an ultra-low-temperature exciton gas, which might result in
an observation of predicted collective phenomena caused by the
high occupation of the lowest energy exciton states. In
addition, the renormalization of the exciton mass due to
in-plane magnetic field was observed; the experimental data are
in a qualitative agreement with the theory.

We thank A. Imamoglu for discussions. We became aware on the
studies of cw PL of interwell excitons in in-plane magnetic
field \cite{Parlangeli}. We are grateful to the authors of Ref.
\cite{Parlangeli} for providing us by their unpublished data and
for discussions. We acknowledge support from INTAS, the Russian
Foundation for Basic Research, and the Programme "Physics of
Solid State Nanostructures" from the Russian Ministry of
Sciences.

\begin{figure}
\caption{
In-plane magnetic field dependence of the time integrated PL
spectrum (a) and the interwell exciton PL kinetics (b) in the
indirect regime ($V_g=1$ V) at $T_{bath}=1.5$ K. Upper inset:
The interwell exciton energy vs in-plane magnetic field, the
line is the fitting curve for small fields
$E_{p=0}=e^2d^2B^2/2Mc^2$, see text. Lower inset: The fastest
interwell exciton PL decay rate and the interwell exciton PL
intensity enhancement after excitation switch off,
$\Delta=ln(I_{PL-max}/I_{PL-pulse-end})$, vs in-plane magnetic
field.
}
\label{fig1}
\end{figure}

\begin{figure}
\caption{
Schematic band diagram of the GaAs/Al$_x$Ga$_{1-x}$As CQW (a).
Scheme of the interwell exciton dispersion at zero (b) and
finite (c) in-plane magnetic field; the photon dispersion as
well as the radiative zone marked bold are also shown.
}
\label{fig2}
\end{figure}

\begin{figure}
\caption{
The interwell exciton PL kinetics vs temperature in in-plane
magnetic field $B=10$ T (a) and at $B=0$ (b). Insets: The
corresponding time integrated PL spectra.
}
\label{fig3}
\end{figure}

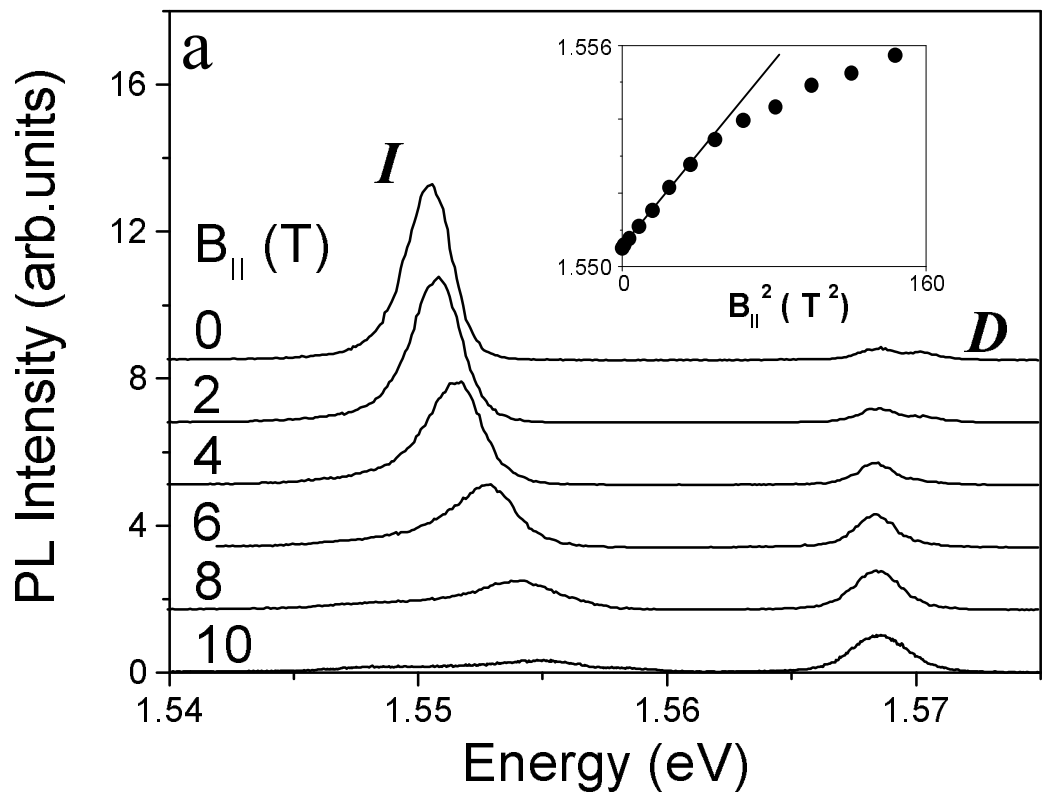
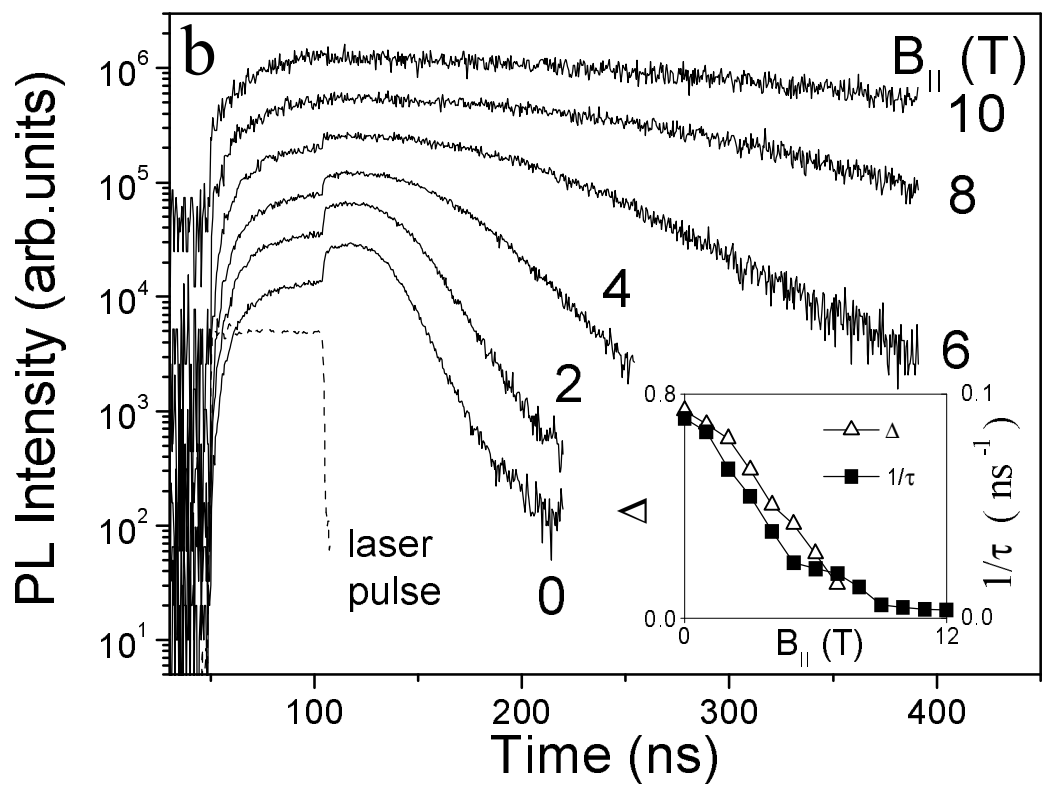
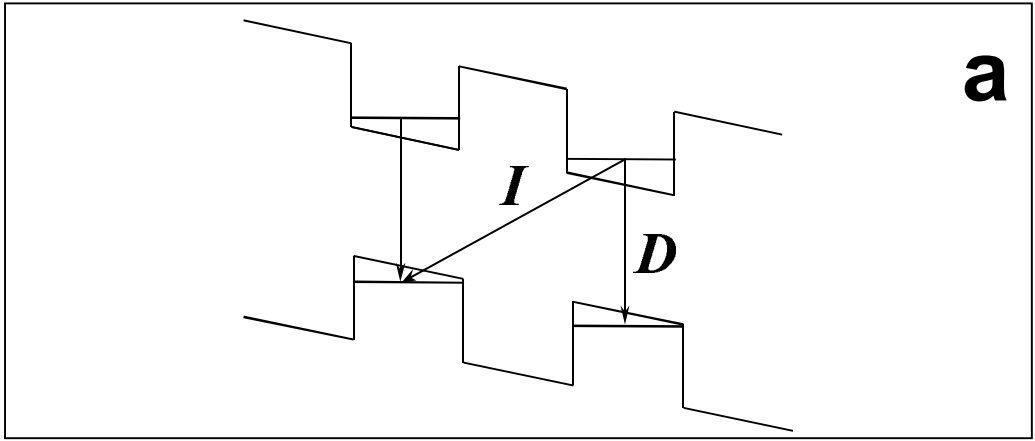
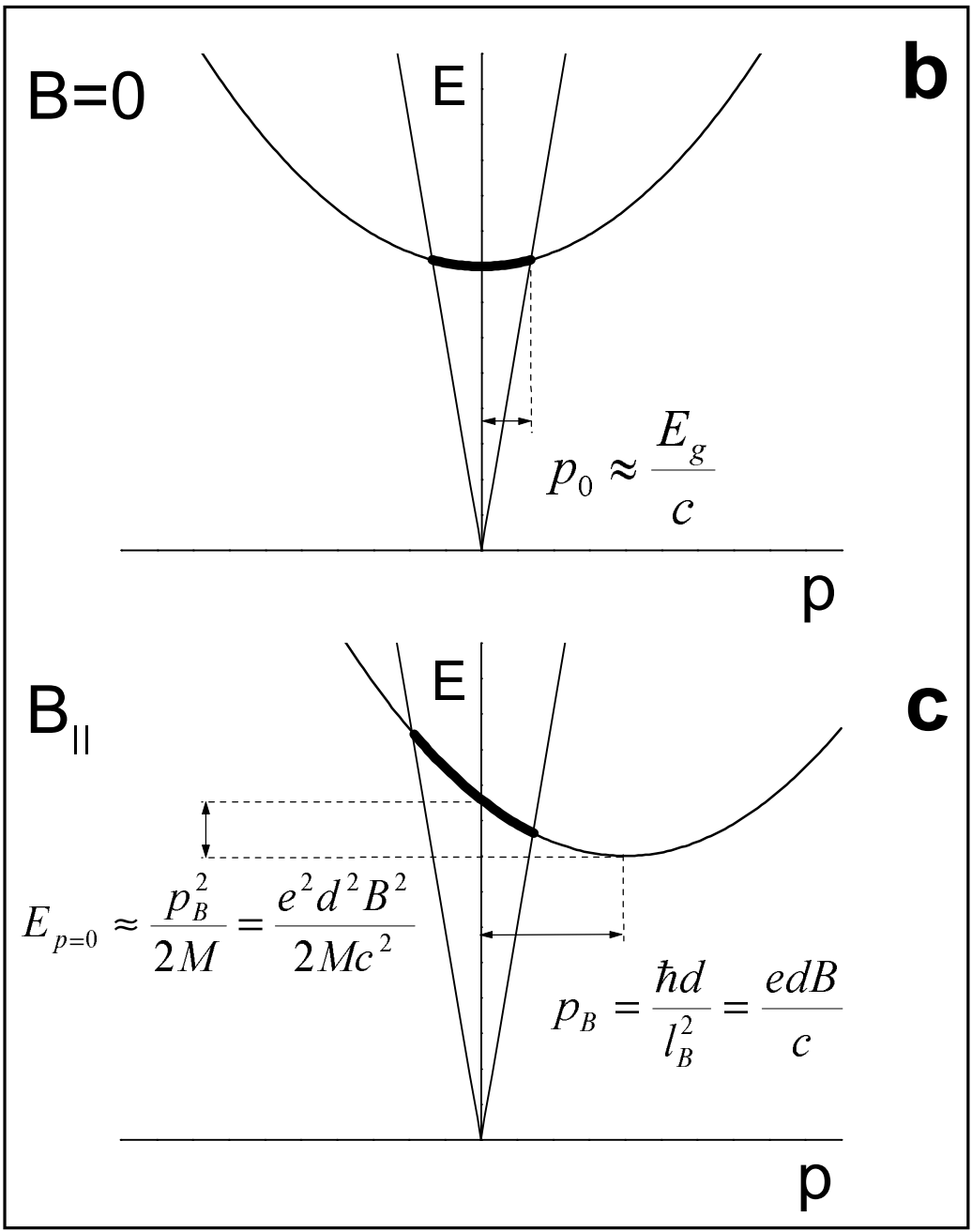
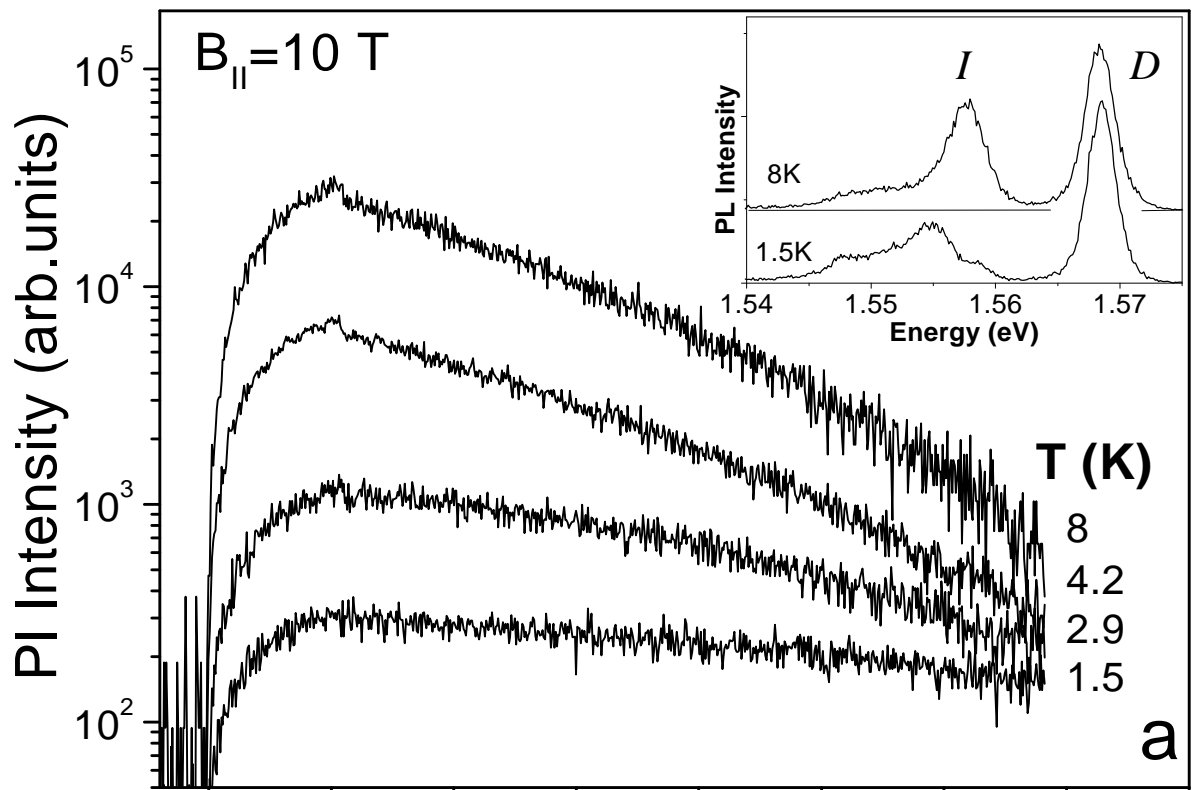
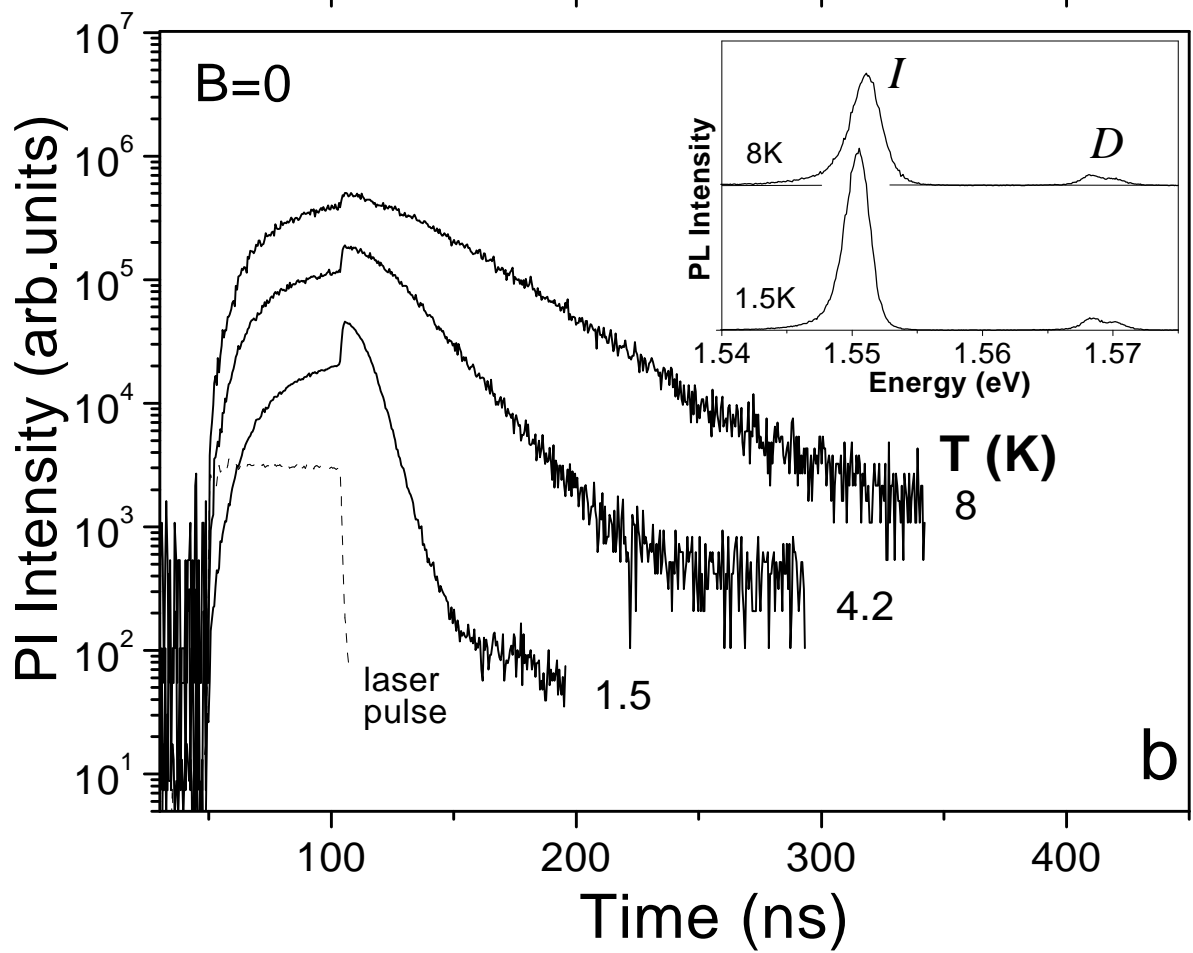
\end{document}